\documentclass[aps,twocolumn,showpacs]{revtex4}
\usepackage[dvips]{graphics}
\usepackage{setspace}
\usepackage{color}
\definecolor{gold}{rgb}{0.85,0.66,0}
\definecolor{dblue}{rgb}{0,0,0.8}

\begin{document}

\title{{\textcolor{gold}{Implementation of Nano-scale Rectifiers: 
An Exact Study}}}

\author{{\textcolor{dblue}{Santanu K. Maiti}}$^{1,2}$}

\affiliation{$^1$Theoretical Condensed Matter Physics Division, 
Saha Institute of Nuclear Physics, 1/AF, Bidhannagar, Kolkata-700 064, 
India \\
$^2$Department of Physics, Narasinha Dutt College, 129 Belilious Road, 
Howrah-711 101, India} 

\begin{abstract}
We propose the possibilities of designing nano-scale rectifiers using
mesoscopic rings. A single mesoscopic ring is used for half-wave 
rectification, while full-wave rectification is achieved using two 
such rings and in both cases each ring is threaded by a time varying 
magnetic flux $\phi$ which plays a central role in the rectification 
action. Within a tight-binding framework, all the calculations are 
done based on the Green's function formalism. We present numerical 
results for the two-terminal conductance and current which support 
the general features of half-wave and full-wave rectifications. The 
analysis may be helpful in fabricating mesoscopic or nano-scale 
rectifiers.
\end{abstract}


\maketitle

\section{Introduction}

Low-dimensional model quantum systems have been the objects of intense
research in theory as well as in experiment since these simple looking
systems are prospective candidates for future generation of nano-devices
in electronic engineering. Several striking features are exhibited
by these systems owing to the quantum interference effect which is
generally preserved throughout the sample only for much smaller sizes, 
while the effect disappears for larger systems. A normal metal 
mesoscopic ring is a very nice example where the electronic motion 
is confined and the transport becomes coherent. Current trend of
miniaturization of electronic devices has resulted much interest in
characterization of ring type nanostructures. There are several
methods for preparation of such rings. For example, gold rings can be
designed by using templates of suitable structure in combination with
metal deposition via ion beam etching~\cite{hobb,pearson}. More recently,
Yan {\it et al.} have prepared gold rings by selective wetting of porous
templates using polymer membranes~\cite{yan}. Using such rings we 
can design nano-scale rectifiers and to reveal this fact the ring is
coupled to two electrodes, to form an electrode-ring-electrode bridge,
where AC signal is applied. Electron transport through a bridge
system was first studied theoretically by Aviram and Ratner~\cite{aviram} 
in $1974$. Following this pioneering work, several experiments have been 
done through different bridge systems to reveal the actual mechanism of 
electron transport. Though, to date a lot of theoretical~\cite{mag,lau,
baer1,baer2,baer3,tagami,gold,cui,orella1,orella2,fowler,peeters} as 
well as experimental works~\cite{reed1,reed2,tali,fish} on two-terminal 
electron transport have been done addressing several important issues, 
yet the complete knowledge of conduction mechanism in nano-scale systems
is still unclear to us. Transport properties are characterized by 
several significant factors like quantization of energy levels, 
quantum interference of electronic waves associated with the geometry 
of bridging system adopts within the junction, etc. Electronic transport 
through a mesoscopic ring is highly sensitive on ring-to-electrode 
interface geometry. Changing the interface structure, transmission 
probability of an electron across the ring can be controlled efficiently. 
Furthermore, electron transport in the ring can also be modulated in 
other way by tuning the magnetic flux, the so-called Aharonov-Bohm (AB) 
flux, penetrated by the ring. The AB flux can change the phases of 
electronic wave functions propagating through different arms of the 
ring leading to constructive or destructive interferences, and 
accordingly, the transmission amplitude changes.

In the present work we illustrate the possibilities of designing nano-scale
rectifiers using mesoscopic rings. To design a half-wave rectifier we use
a single mesoscopic ring, while two such rings are considered for designing 
a full-wave rectifier. Both for the cases of half-wave and full-wave 
rectifiers, each ring is threaded by a time varying magnetic flux $\phi$ 
which plays the central role for the rectification action. Within a 
tight-binding framework, a simple parametric approach~\cite{muj1,san3,muj2,
san1,sam,san2,hjo,walc1,walc2} is given and all the calculations are 
done through single particle Green's function technique to reveal the 
electron transport. Here we present numerical results for the two-terminal
conductance and current which clearly describe the conventional features
of half-wave and full-wave rectifications. Our exact analysis may be 
helpful for designing mesoscopic or nano-scale rectifiers. To the best 
of our knowledge the rectification action using such simple mesoscopic 
rings has not been addressed earlier in the literature.

The paper is organized as follows. With the brief introduction 
(Section I), in Section II, we describe the model and the theoretical 
formulations for our calculations. Section III presents the significant 
results, and finally, we conclude our results in Section IV.

\section{Model and synopsis of the theoretical background}

\subsection{Circuit configuration of a half-wave rectifier}

\begin{figure}[ht]
{\centering \resizebox*{7.5cm}{3.5cm}{\includegraphics{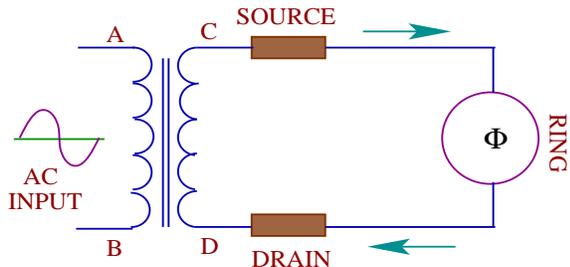}}\par}
\caption{(Color online). Circuit configuration for a half-wave rectifier,
where a mesoscopic ring, subject to a time varying magnetic flux $\phi$, 
is attached symmetrically to source and drain. These electrodes are directly
coupled to the end points C and D of secondary winding of the transformer.
The arrow indicates current direction in the circuit.}
\label{halfwave}
\end{figure}

We begin by referring to Fig.~\ref{halfwave}. A mesoscopic ring, threaded
by a time varying magnetic flux $\phi$, is attached symmetrically to two
semi-infinite one-dimensional ($1$D) metallic electrodes, namely, source 
and drain. Two end points C and D of secondary winding of the transformer
are directly coupled to the source and drain, respectively, through which
AC signal is applied. Mathematically, we can express the AC signal in the
form,
\begin{equation}
V(t)=V_0 \sin(\omega t)
\label{ac}
\end{equation}
where, $V_0$ is the peak voltage, $\omega$ corresponds to the angular
frequency and $t$ represents the time. For the rectification action, we
\begin{figure}[ht]
{\centering \resizebox*{7.5cm}{4.5cm}{\includegraphics{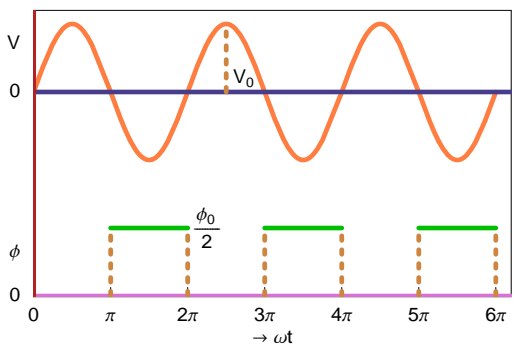}}\par}
\caption{(Color online). Variations of AC input signal (orange line) and
magnetic flux (green line) as a function of time $t$.}
\label{signal1}
\end{figure}
choose the flux passing through the ring in the square wave form as a
function of time. Mathematically it can be written as,
\begin{eqnarray}
\phi(t) & = & 0 ~~~~~~{\mbox{for}}~~~n\pi<\omega t <(n+1) \pi \nonumber \\
 & & ~~~~{\mbox{where,}}~n=0 ~{\mbox{or any even integer}} \nonumber \\
 & = & \frac{\phi_0}{2} ~~~~{\mbox{for}}~~~n\pi<\omega t <(n+1) \pi 
\nonumber \\
 & & ~~~~{\mbox{where,}}~n= {\mbox{any odd integer}} 
\label{flux}
\end{eqnarray} 
where, $\phi_0$ ($=ch/e$) is the elementary flux-quantum. Both the AC
signal and threaded magnetic flux $\phi$ vary periodically with the
same frequency $\omega$. The variations of AC input signal (orange line)
and magnetic flux (green line) are represented graphically in 
Fig.~\ref{signal1}. From the spectra it is clear that in the negative
half-cycles of the input signal, the ring is penetrated by the flux
$\phi_0/2$, which is required for half-wave rectification.

\subsection{Circuit configuration of a full-wave rectifier}

\begin{figure}[ht]
{\centering \resizebox*{7.5cm}{5cm}{\includegraphics{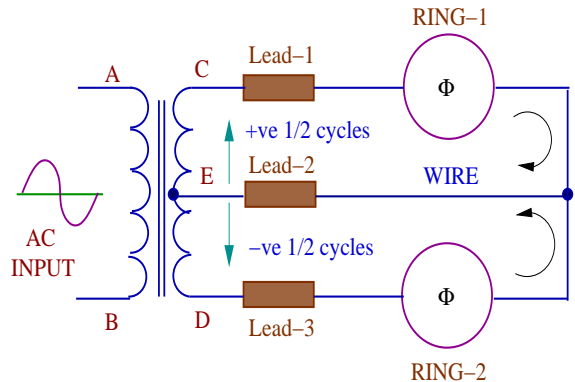}}\par}
\caption{(Color online). Circuit configuration for a full-wave rectifier
using two mesoscopic rings, where each ring is threaded by a time varying
magnetic flux. The arrows indicate current directions in the circuit.}
\label{fullwave}
\end{figure}

The circuit diagram for a full-wave rectifier is schematically presented
in Fig.~\ref{fullwave}. Here two mesoscopic rings are used, together with
a transformer whose secondary winding is split equally into two and has a 
common center tapped connection, described by the point E. The rings are
threaded by time varying magnetic fluxes $\phi_1$ and $\phi_2$, 
\begin{figure}[ht]
{\centering \resizebox*{7.5cm}{5cm}{\includegraphics{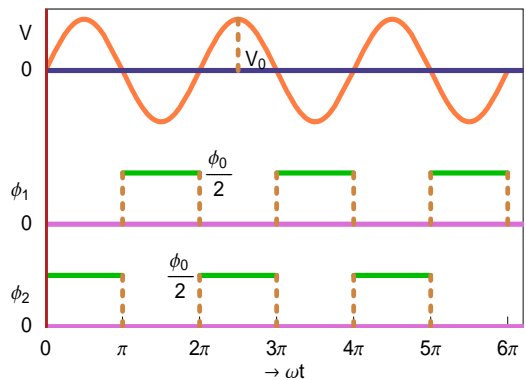}}\par}
\caption{(Color online). Variations of AC input voltage (orange curve)
and magnetic fluxes (green curves) as a function of time $t$. A constant 
$\pi$ phase shift exists between the fluxes, $\phi_1$ and $\phi_2$, 
threaded by the rings.}
\label{signal2}
\end{figure}
respectively, and their functional forms are described by 
Eq.~\ref{flux}. A constant $\pi$ phase shift exists between these two
fluxes, as schematically shown in Fig.~\ref{signal2}, which is essential 
for full-wave rectification. In the circuit there are three electrodes,
namely, lead-1, lead-2 and lead-3, those are directly coupled to the
points C, E and D of secondary winding of the transformer, respectively.
During positive half-cycle of the AC input signal, lead-1 and lead-2
act as source and drain, respectively. While, for the negative half-cycle
of the AC input voltage, lead-3 and lead-2 behave as source and drain,
respectively.
Thus the circuit consists of two half-wave rectifiers, as shown clearly 
from Fig.~\ref{fullwave}. Each rectifier is connected to a single quantum 
wire through which current flows in one direction during each half-cycle of 
the AC input signal which is described by the arrows in Fig.~\ref{fullwave}. 
The variations of AC input voltage (orange curve) and associated magnetic 
fluxes (green curves) in the rings are represented graphically in 
Fig.~\ref{signal2}. From the spectra it is observed that, in the positive 
half-cycles of AC input signal, ring-2 is threaded by the flux $\phi_0/2$. 
On the other hand, for the negative half-cycles of the input signal, 
ring-1 is penetrated by the flux $\phi_0/2$.

\subsection{Theoretical formulation}

Using Landauer conductance formula~\cite{datta,marc} we determine
two-terminal conductance ($g$) of the mesoscopic ring. At much low 
temperatures and bias voltage it ($g$) can be written in the form,
\begin{equation}
g=\frac{2e^2}{h} T
\label{equ1}
\end{equation}
where, $T$ corresponds to the transmission probability of an electron 
across the ring. In terms of the Green's function of the ring and 
its coupling to two electrodes, the transmission probability can be 
expressed as~\cite{datta,marc},
\begin{equation}
T={\mbox{Tr}} \left[\Gamma_1 G_{R}^r \Gamma_2 G_{R}^a\right]
\label{equ2}
\end{equation}
where, $\Gamma_S$ and $\Gamma_D$ describe the coupling of the ring 
to the source and drain, respectively. Here, $G_R^r$ and $G_R^a$ 
are the retarded and advanced Green's functions, respectively, of the 
ring considering the effects of the electrodes. Now, for the full system 
i.e., the mesoscopic ring, source and drain, the Green's function is 
expressed as,
\begin{equation}
G=\left(E-H\right)^{-1}
\label{equ3}
\end{equation}
where, $E$ is the energy of the source electron. Evaluation of this 
Green's function needs the inversion of an infinite matrix, which is 
really a difficult task, since the full system consists of the finite 
size ring and two semi-infinite $1$D electrodes. However, the full 
system can be partitioned into sub-matrices corresponding to the 
individual sub-systems and the effective Green's function for the 
ring can be written in the form~\cite{marc,datta},
\begin{equation}
G_R=\left(E-H_R-\Sigma_S-\Sigma_D \right)^{-1}
\label{equ4}
\end{equation}
where, $H_R$ describes the Hamiltonian of the ring. Within the 
non-interacting picture, the tight-binding Hamiltonian of the ring 
can be expressed like,
\begin{equation}
H_R = \sum_i \epsilon_i c_i^{\dagger} c_i + \sum_{<ij>} v 
\left(c_i^{\dagger} c_j e^{i\theta}+ c_j^{\dagger} c_i e^{-i\theta}\right)
\label{equ5}
\end{equation}
where, $\epsilon_i$ and $v$ correspond to the site energy and 
nearest-neighbor hopping strength, respectively. $c_i^{\dagger}$ ($c_i$) 
is the creation (annihilation) operator of an electron at the site $i$
and $\theta=2 \pi \phi/N \phi_0$ is the phase factor due to the flux 
$\phi$ enclosed by the ring consists of $N$ atomic sites. A similar 
kind of tight-binding Hamiltonian is also used, except the phase factor 
$\theta$, to describe the electrodes where the Hamiltonian is parametrized 
by constant on-site potential $\epsilon^{\prime}$ and nearest-neighbor 
hopping integral $t^{\prime}$. The hopping integral between the ring and
source is $\tau_S$, while it is $\tau_D$ between the ring and drain. 
In Eq.~(\ref{equ4}), $\Sigma_S$ and $\Sigma_D$ are the self-energies 
due to the coupling of the ring to the source and drain, respectively, 
where all the information of the coupling are included into these 
self-energies.

To determine current, passing through the mesoscopic ring,
we use the expression~\cite{marc,datta},
\begin{equation}
I(V)=\frac{2 e}{h}\int \limits_{-\infty}^{\infty} 
\left(f_S-f_D\right) T(E)~ dE
\label{equ6}
\end{equation}
where, $f_{S(D)}=f\left(E-\mu_{S(D)}\right)$ gives the Fermi distribution
function with the electrochemical potential $\mu_{S(D)}=E_F\pm eV/2$ and
$E_F$ is the equilibrium Fermi energy. For the sake of simplicity,
we take the unit $c=e=h=1$ in our present calculations. 

\section{Numerical results and discussion}

To illustrate the results, let us begin our discussion by mentioning the 
values of different parameters used for our numerical calculations. 
In the mesoscopic ring, the on-site energy $\epsilon_i$ is fixed to $0$ 
for all the atomic sites $i$ and nearest-neighbor hopping strength 
$v$ is set to $3$. While, for the side-attached electrodes the on-site 
energy ($\epsilon^{\prime}$) and nearest-neighbor hopping strength 
($t^{\prime}$) are chosen as $0$ and $4$, respectively. The hopping 
strengths $\tau_S$ and $\tau_D$ are set as $\tau_S=\tau_D=2.5$. The 
equilibrium Fermi energy $E_F$ is taken as $0$.

\subsection {Half-wave rectification}

\begin{figure}[ht]
{\centering \resizebox*{7.5cm}{4.75cm}{\includegraphics{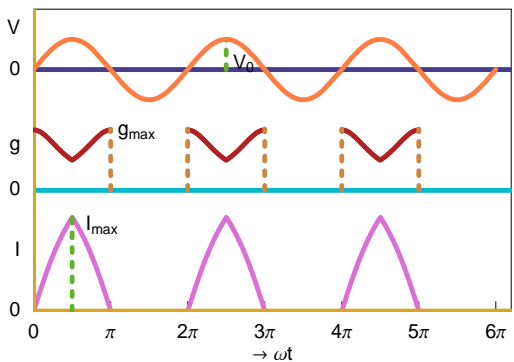}}\par}
\caption{(Color online). Half-wave rectification action. Upper panel
describes AC input signal of amplitude $V_0$. Middle panel illustrates
the variation of conductance $g$ with $\omega t$. Lower panel represents
the variation of current $I$ as a function of $\omega t$. $g_{max}$ and
$I_{max}$ correspond to the conductance and current maxima, respectively.
Here we set the ring size $N=8$.}
\label{current1}
\end{figure}

The rectification action of the half-wave rectifier is illustrated
in Fig.~\ref{current1}. In the upper panel of this figure we plot AC input 
signal (orange curve) as a function of $\omega t$. The amplitude $V_0$ of 
the AC signal is fixed at $2$. The variation of conductance $g$ (red curve) 
as a function of $\omega t$ is shown in the middle panel. From the
results we clearly observe that only in the positive half-cycles of 
the AC input signal, conductance exhibits finite value. On the other
hand, conductance exactly drops to zero for the negative half-cycles
of the AC input signal. $g_{max}$ is the amplitude of the conductance 
which gets the value $2$, and therefore, the transmission probability
$T$ goes to unity, according to Landauer conductance formula (see
Eq.~\ref{equ1}). Now we try to figure out the rectification operation.
The probability amplitude of getting an electron from the source to
drain across the ring depends on the quantum interference effect of 
the electronic waves passing through the upper and lower arms of the 
ring. For a symmetrically connected ring (upper and lower arms are 
identical to each other), threaded by a magnetic flux $\phi$, the 
probability amplitude of getting an electron across the ring becomes 
exactly zero ($T=0$) for the typical flux, $\phi=\phi_0/2$. This 
vanishing behavior of the transmission probability can be easily 
obtained through few simple mathematical steps. Therefore, in the 
negative half cycles of the AC input signal, electron conduction 
through the ring is no longer possible since the ring encloses the
flux $\phi_0/2$ for these half-cycles. Only for the positive
half-cycles of the AC input signal electron can conduct through
the ring as the ring is not penetrated by the flux $\phi_0/2$.
This feature clearly describes the half-wave rectification action.
To visualize the rectification feature more prominently we present 
the variation of current (magenta curve) as a function of $\omega t$ 
in the lower panel of Fig.~\ref{current1}. We obtain the current 
$I$ through the ring by integrating over the transmission function
$T$ (see Eq.~\ref{equ6}). Following the conductance pattern (red
curve), the vanishing nature of the current in the negative 
half-cycles of the AC input signal is clearly understood. The
non-vanishing behavior (magenta curve) of the current is only achieved 
for the positive half-cycles of the AC input. All these features clearly
support the half-wave rectification action as we get in traditional
macroscopic half-wave rectifiers.

\subsection {Full-wave rectification}

\begin{figure}[ht]
{\centering \resizebox*{7.5cm}{4.75cm}{\includegraphics{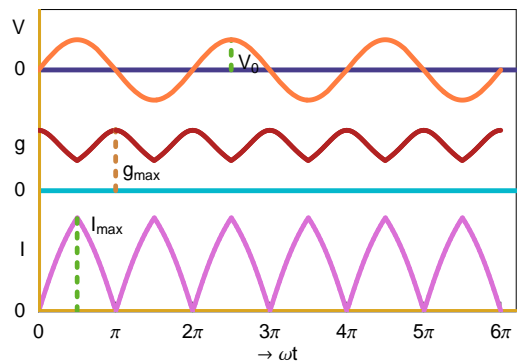}}\par}
\caption{(Color online). Full-wave rectification action. Upper panel
describes AC input signal of amplitude $V_0$. Middle panel illustrates
the variation of conductance $g$ with $\omega t$. Lower panel represents
the variation of current $I$ as a function of $\omega t$. $g_{max}$ and
$I_{max}$ correspond to the conductance and current maxima, respectively.
Here we fix the ring size $N=8$.}
\label{current2}
\end{figure}

Next, we investigate the behavior of full-wave rectification mechanism 
as illustrated in Fig.~\ref{current2}. The upper, middle and lower
panels correspond to the time varying features of input signal,
conductance and current, respectively. Our results demonstrate that both 
for the positive and negative half-cycles of the input signal non-zero 
conductance (red curve) is obtained. During the positive half-cycles 
ring-1 conducts, while for the negative half-cycles conduction takes
place through ring-2, as both the two rings enclose time varying
magnetic fluxes having identical magnitude $\phi_0/2$ with a constant 
phase shift $\pi$. The feature of full-wave rectification i.e., 
obtaining non-zero conductance for both positive and negative 
half-cycles of the input signal is attributed to the quantum 
interference effect as explained earlier in case of half-wave 
rectification. In the same footing, here we describe the variation 
of current $I$ (magenta curve) with time $t$ to support the full-wave 
rectification operation properly. Here also non-zero value of current
is achieved both for the positive and negative half-cycles of the
input signal following the conductance spectrum. 
Our results support the full-wave rectification action and agree well 
with the basic features obtained in conventional macroscopic full-wave
rectifiers.

\section{Concluding remarks}

In a nutshell, we have proposed the possibilities of designing nano-scale
rectifiers using mesoscopic rings enclosing a time varying magnetic flux.
The half-wave rectifier is designed using a single mesoscopic rings,
while two such rings are used for full-wave rectification. The rectification
action is achieved using the central idea of quantum interference effect
in presence of flux $\phi$ in ring shaped geometries. We adopt a simple
tight-binding framework to illustrate the model and all the calculations
are done using single particle Green's function formalism. Our exact
numerical results provide two-terminal conductance and current which
support the general features of half-wave and full-wave rectifications.
Our analysis can be used in designing tailor made nano-scale rectifiers.

Throughout our work, we have addressed all the essential features
of rectification operation considering a ring with total number of 
atomic sites $N=8$. In our model calculations, this typical
number ($N=8$) is chosen only for the sake of simplicity. Though the
results presented here change numerically with the ring size ($N$), but
all the basic features remain exactly invariant. To be more specific, it
is important to note that, in real situation the experimentally
achievable rings have typical diameters within the range $0.4$-$0.6$
$\mu$m. In such a small ring, unrealistically very high magnetic fields
are required to produce a quantum flux. To overcome this situation,
Hod {\em et al.} have studied extensively and proposed how to construct
nanometer scale devices, based on Aharonov-Bohm interferometry, those
can be operated in moderate magnetic fields~\cite{baer4,baer5,baer6,baer7}.

In the present paper we have done all the calculations by ignoring
the effects of the temperature, electron-electron correlation, etc. 
Due to these factors, any scattering process that appears in the
mesoscopic ring would have influence on electronic phases, and, in
consequences can disturb the quantum interference effects. Here we
have assumed that, in our sample all these effects are too small, and
accordingly, we have neglected all these factors in this particular 
study.

The importance of this article is mainly concerned with (i) the simplicity 
of the geometry and (ii) the smallness of the size.

\end{document}